\documentclass[10pt]{iopart}
\usepackage{graphicx}
\usepackage[amssymb]{SIunits}

\begin{document}

\title[Light / heavy hole switching with correlated characterization on a single quantum dot]{Probing the light hole / heavy hole switching with correlated magneto-optical spectroscopy and chemical analysis on a single quantum dot}

\author{A Artioli$^{1,2}$, P Rueda-Fonseca$^{1,2}\footnote{Present
address: Aledia, CEA/MINATEC campus, 38000 Grenoble, France.}$, K
Moratis$^{1,2}$, J F Motte$^1$, F Donatini$^1$, M den Hertog$^1$, E
Robin$^2$, R Andr\'{e}$^1$, Y-M Niquet$^2$, E Bellet-Amalric$^2$, J
Cibert$^1$, and D Ferrand$^1$}

\address{$^1$ Universit\'{e} Grenoble-Alpes, CNRS, Institut N\'eel, 38000 Grenoble, France}
\address{$^2$ Universit\'{e} Grenoble-Alpes, CEA, INAC, 38000 Grenoble, France}

\ead{joel.cibert@neel.cnrs.fr}

\vspace{10pt}
\begin{indented}
\item[]November 2018
\end{indented}

\begin{abstract}
A whole series of complementary studies have been performed on the
same, single nanowire containing a quantum dot: cathodoluminescence
spectroscopy and imaging, micro-photoluminescence spectroscopy under
magnetic field and as a function of temperature, and
energy-dispersive X-ray spectrometry and imaging. The ZnTe nanowire
was deposited on a Si$_3$N$_4$ membrane with Ti/Al patterns. The
complete set of data shows that the CdTe quantum dot features the
heavy-hole state as a ground state, although the compressive
mismatch strain promotes a light-hole ground state as soon as the
aspect ratio is larger than unity (elongated dot). A numerical
calculation of the whole structure shows that the transition from
the heavy-hole to the light-hole configuration is pushed toward
values of the aspect ratio much larger than unity by the presence of
a (Zn,Mg)Te shell, and that the effect is further enhanced by a
small valence band offset between the semiconductors in the dot and
around it.

\end{abstract}

\vspace{2pc} \noindent{\it Keywords}: nanowires, quantum dot,
semiconductors, molecular beam epitaxy, optical spectroscopy,
cathodoluminescence, EDX

%
% For two-column output uncomment the next line and choose [10pt] rather than [12pt] in the \documentclass declaration
\ioptwocol

\section{Introduction}
Quantum confinement in semiconductor nanostructures is currently
explored in order to achieve specific functions in the field of
nanophotonics, nanoelectronics, nanomagnetism, quantum information,
communication and sensing. A prerequisite is to be able to address a
single object, for instance to achieve the emission of single
photons or entangled photon pairs, or various realizations of
qubits. Addressing and scaling constitute a clear advantage of
semiconductors with respect to other systems; however there is a
large dispersion from one single object to another one, up to the
point that in order to reach a better understanding of the relevant
physical mechanisms, it is mandatory to apply complementary
experimental methods not to an ensemble nor to a collection of
single objects, but all to the very same single object. Here we
present and exploit a multiple-technique approach which can be
applied to any system which can be physically manipulated, such as a
nanowire (NW), a quantum dot (QD) in a NW, a nanocristallite, or any
system where the actual configuration controls the electronic,
photonic and spin properties.

The system under study is a QD in a NW. The advantage of fabricating
a QD insertion in a NW is that it permits to combine a large variety
of different materials and to control the size and shape of the QD
over a very broad range, much broader than by using the usual
Stranski-Krastanow (SK) growth mode. It is thus possible to design
the aspect ratio and mismatch strain configurations, in such a way
that the ground state at the top of the valence band is the
light-hole state \cite{Niquet, Zielinski,Jeannin}. A proper control
of the hole state has decisive consequences on many physical
properties, such as the selection rules of excitons and charged
excitons and the opportunity to optically manipulate the qubit
formed on a confined carrier, the diagram of light emission and
polarization, the spin anisotropy of a confined carrier and the
corresponding dynamics, or the magnetic anisotropy in case of
coupling with magnetic impurities, to cite but a few. The selection
rules for an exciton associated with a light hole naturally form the
so-called lambda-configuration required for the manipulation of the
spin state of a confined carrier: the solution exploited so far
\cite{Huo} involves complex structures incorporating a ferroelectric
stressor, while using a mismatched elongated QD offers a
straightforward solution. By contrast, the orientation of the
emitting dipole realized by a heavy hole exciton is more adapted to
light guiding: the low emission efficiency of InAsP/InP QDs in NWs
was attributed \cite{Haffouz} to the use of elongated QDs in order
to shift the wavelength to the telecom band (1.55 $\mu$m), thus
favoring a light-hole exciton. The design of the shell is also of
prime importance, including tapering \cite{Haffouz} to improve the
coupling to a photonic structure or free space propagation, and an
asymmetric cross section \cite{Foster, Schlehahn} to drive the
emission into a single linearly polarized optical mode: in all
cases, the control of the shape and size of the QD and its position
within a NW with a well controlled profile is needed, and we
demonstrate here that the EDX tomographic approach \cite{RuedaEDX}
can be combined with a spectroscopic and photonic characterization.
Finally, the design of the spin anisotropy of the confined carrier
has many consequences: an example is the fine structure and the
associated spin dynamics of the carrier contemplated as a qubit.
Another interesting aspect is the magnetic anisotropy of the
magnetic polaron formed in a dot made of a dilute magnetic
semiconductor such as Cd$_{1-x}$Mn$_x$Te. While magnetic polarons
formed around a heavy-hole exciton have been studied for a long time
\cite{Klopo2011}, they were not yet observed in structures where
light holes could be favored \cite{Szymura, Plachta}. Finally,
combining chemical analysis and spectroscopy on the same nanoobject
would be highly relevant in the case of colloidal nanoparticles: to
stay with dilute magnetic system, we just mention II-VI
nanoparticles \cite{Delikanli,Muckel} and halide perovskites
\cite{Guria, Mir}.

In the simple case of a QD made of an isotropic material embedded in
a thick matrix with a smaller lattice parameter, the built-in
elastic strain (Eshelby model \cite{Eshelby}) tends to link the
ground state to the heavy-hole in a flat ellipsoidal QD, and to the
light-hole in an elongated QD; the boundary is the spherical QD
where the fourfold degeneracy is restored. The Luttinger Hamiltonian
and the anisotropic elastic properties should be taken into account
but both feature cubic symmetry and thus should not lift this
fourfold degeneracy. However an additional shell with a different
lattice parameter is often added in order to passivate the surface
and it may significantly affect the position of the
light-hole/heavy-hole crossover.

This is the case of the QD under study, a Cd$_{1-x}$Mn$_x$Te
elongated insertion embedded in a ZnTe/Zn$_{1-y}$Mg$_y$Te core-shell
NW: the ZnTe core has the smallest lattice parameter, so that the
elongated dot favors the light hole while the outer shell tends to
restore the heavy hole. Most QDs from the same sample exhibit a
light-hole ground state (as demonstrated in our previous study
\cite{Jeannin}) but some exhibit a heavy-hole one. Our
multiple-technique approach combines (1) high-resolution scanning
transmission electron microscopy combined with energy-dispersive
x-ray spectrometry (EDX) with imaging and 3D reconstruction through
discrete tomography, (2) scanning electron microscopy (SEM) with
low-temperature cathodoluminescence (CL) spectroscopy and imaging,
and (3) micro-photoluminescence (PL) at variable temperature under
applied magnetic field. All these experimental methods were applied
to the same QD, placed on a dedicated micro-fabricated silicon
nitride membrane using a micromanipulator. We show that this QD
hosts a heavy-hole ground state in spite of its elongated shape with
an aspect ratio definitely larger than unity; a calculation of the
hole states by using the TB$\_$Sim package of codes \cite{code}
shows that the strain induced by the shell pushes the heavy-hole /
light-hole crossing to values of the aspect ratio larger than unity,
the effect being enhanced if the valence band offset is small.
\section{Methods}
\begin{figure}
\centering
\includegraphics [scale=1]{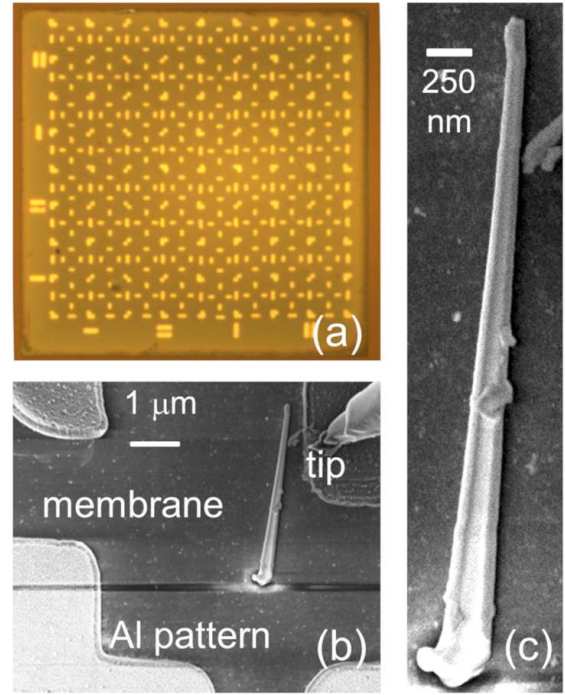}
\caption{(a) Optical image of the Si$_3$N$_4$ patterned membrane;
(b) SEM image showing the NW deposited on the membrane (the tip of
the nanomanipulator is visible in the top-right corner); (c) SEM
image zoom of the NW.} \label{fig1}
\end{figure}

The micro-fabricated TEM grid $/$ PL sample holder is based on
Si$_3$N$_4$ membranes, thin enough (50 nm) to be transparent to an
electron beam, and compatible with low-temperature spectroscopy. The
same type of membrane was used to identify the influence of crystal
structure and defects on electronic properties, for instance to
determine the influence of the polarity of a contacted GaN-AlN NW
heterostructure on its current-voltage characteristics
\cite{Hertog2012}, or to associate a specific PL line to polytypes
\cite{Bao,Senichev} or polarity inversion boundaries \cite{Auzelle}.
Here we go one step further and combine a quantitative study of the
morphology of a nanofabricated single object (the composition
profile of a QD embedded in a NW) and the study of its electronic
properties. The silicon nitride membranes were fabricated from a 300
$\mu$m thick Si(100) four-inch wafer with 50 nm of stoichiometric
Si$_3$N$_4$ deposited by LioniX International on both sides. Using
laser lithography followed by a reactive ion etching step a
pattern is written in the Si$_3$N$_4$ layer. Then a KOH bath at
80$^\circ$C etches through the Si on the other side of the wafer,
leaving only the layer of Si$_3$N$_4$, which now constitutes the
membrane. Next, markers are defined by UV laser lithography and
electron beam deposition of Ti/Al (10 nm / 50 nm), followed by a
lift-off step to remove the metal deposited on the resist; these
markers (Fig.~\ref{fig1}a) allow us to rapidly identify the NW all
along the chain of experimental set-ups.

%insertion figure

The sample was grown as described previously \cite{RuedaGrowth}. We
used a (111)B Cd$_{0.96}$Zn$_{0.04}$Te substrate with a small
misorientation (\unit{2}{\degree}) to avoid the formation of twins
during the growth of the preliminary ZnTe buffer layer. The NW (NW
S2 of our EDX study \cite{RuedaEDX}) was transferred from the
as-grown sample to the sample holder $/$ silicon nitride membrane,
using the micromanipulator of a dual beam FIB/SEM machine (Zeiss
NVision 40 DualBeam), see Fig.~\ref{fig1}b.

The CL set-up consists of a FESEM FEI Inspect F50 to ensure a high
spatial resolution (a few nm at 5 keV), with a IHR550 spectrometer
equipped with a CL accessory and a low-temperature Gatan stage to
cool the sample down to $6$~K. It is often convenient to perform CL
as a first characterization, implying that the electron beam
intensity and voltage be kept low enough, typically 250 pA and 5keV,
to reduce defect formation.

The magneto-optical properties were studied using a low temperature
confocal micro-PL set-up \cite{Artioli2013}. The sample holder was
mounted in a helium-flow cryostat (Oxford Spectromag), with
temperature between 300~K and 4.2~K, and a magnetic field up to 11~T
in the Faraday configuration (excitation and detection along the
magnetic field, perpendicular to the NW axis). The NW was excited by
a 488~nm cw laser beam focused to 4~$\mu$m$^{2}$, using a 100X
Mitutoyo magnification microscope and a piezoelectric scanner. Unless mentioned, the laser power was kept well below 1~$\mu$W in order to minimize heating. The
PL signal was collected by the same microscope objective and sent
into a 0.46~m Jobin-Yvon spectrometer equipped with a CCD camera.

EDX spectrometry was performed in a FEI Tecnai Osiris S/TEM equipped
with four Silicon Drift Detectors, operated at 200 keV. The EDX
signal is collected as an hypermap where each pixel corresponds to
the x-ray emission spectrum of atoms along the electron beam. We
used the Quantax-800 software from Bruker for background correction
and deconvolution, to extract the contributions of the L lines of
Te, Cd, Au, and K lines of Zn, Mg, and O.  A transparent sample
holder must be used if the goal is a quantitative analysis of the
EDX in order to extract chemical profiles, as the signal from a bulk
sample holder would totally blind the detectors. If needed, it is
possible to rotate the NW around its axis: discrete tomography is
then achieved by recording at least 2 projections along two nearly
orthogonal axes \cite{RuedaEDX}.

The structural and electronic properties of nanowires were
calculated numerically with the TB$\_$Sim code \cite{code}. The
strains are first computed with a finite elements discretization of
continuous elasticity equations. The Poisson equation is then solved
for the resulting piezoelectric potential. Finally, the hole states
are calculated with a six-bands \textbf{k.p} model discretized on
the same mesh. The lattice parameters, elastic, dielectric and
piezoelectric constants of the materials are: $a = $6.481 {\AA},
$c_{11} = $61.5 GPa, $c_{12} = $43 GPa, $c_{44} = $19.6 GPa,
$\varepsilon_r = $10.6, $e_{14} = $0.03 C~m$^{-2}$ for CdTe, and $a
= $6.104 {\AA}, $c_{11} = $71.6 GPa, $c_{12} = $40.7 GPa, $c_{44} =
$31.2 GPa, $\varepsilon_r = $10.1, $e_{14} = $0.03 C~m$^{-2}$ for
ZnTe. The Luttinger parameters, spin-orbit energy and deformation
potentials are: $\gamma_1 = $4.6, $\gamma_2 = $1.6, $\gamma_3 =
$1.8, $\Delta = $0.9 eV, $a_v = $0.55 eV, $b = $-1.23 eV, $d = $-5.1
eV for CdTe and $\gamma_1 = $4.07, $\gamma_2 = $0.78, $\gamma_3 =
$1.59, $\Delta = $ 0.95 eV, $a_v = $0.79 eV, $b = -$1.3 eV, $d =
-$4.3 eV for ZnTe. The parameters of (Zn,Mg)Te are the same as those
of ZnTe, except $a = $6.152 {\AA}. We vary the band offset between
CdTe and ZnTe. The bandgap of MgTe is 1~eV larger than that of ZnTe,
with approximately $1/3$ in the valence band. Note that the valence
band offset of the external shell is expected to have no effect on
the hole states which are localized in the dot or at its interface.

\section{Results}

%insertion figure
\begin{figure}
\centering
\includegraphics*[scale=1]{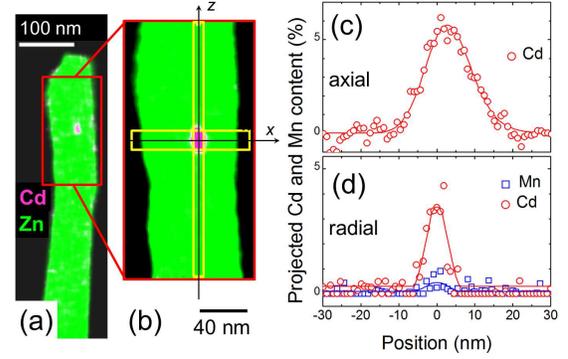}
\caption{(a) EDX image of the nanowire tip; (b) EDX image zoom
around the QD showing the Zn-rich and Cd-rich areas; (c-d) Cd and Mn
chemical profiles along the two axes (axial$=z$ and radial$=x$)
defined in (b); symbols display the experimental data and solid lines show the fits described in text.} \label{fig1b}
\end{figure}

Figure \ref{fig1}c shows the SEM image of the NW, and
Fig.~\ref{fig1b}a is an EDX image of the NW tip, showing the areas
which are predominantly ZnTe and CdTe. Such images are qualitative.
A quantitative analysis \cite{RuedaEDX} is based on the intensity
profiles across the QD, along the two perpendicular axes,
Fig.~\ref{fig1b}c-d. A good fit of the radial intensity profile,
Fig.~\ref{fig1b}d, was obtained \cite{RuedaEDX} assuming a uniform
composition, Cd$_{0.9}$Mn$_{0.1}$Te, in a cylinder of diameter 8~nm.
The fit of the axial profile, Fig.~\ref{fig1b}c, is obtained
using the expression proposed earlier\cite{Orru}: a square profile
of width $L$, broadened by an exponential of characteristic size
$\tau$ (reservoir effect) and a Gaussian of width $\sigma$, with
$L=12$~nm, $\tau$=3.5~nm and $\sigma$=5~nm. Other radial profiles
have been measured and analyzed for several positions along the NW
axis \cite{RuedaEDX}: in addition to the Cd$_{0.9}$Mn$_{0.1}$Te QD,
they clearly identify the (Zn,Mg)Te external shell, but also a very
thin (Zn,Cd)Te internal shell due to lateral growth.

\begin{figure}
\centering
\includegraphics[scale=1]{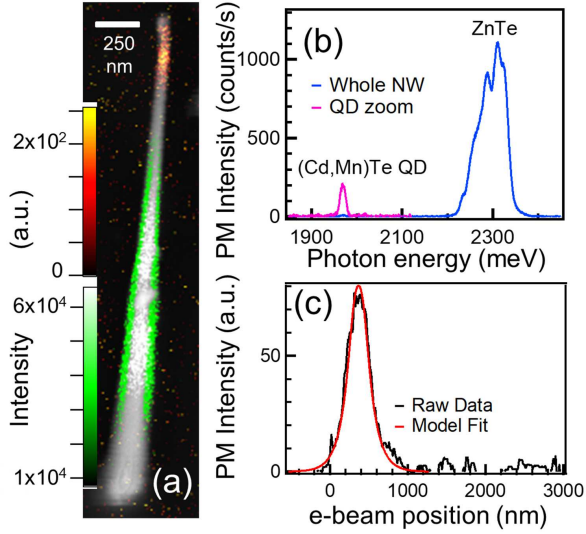}
\caption{(a) CL monochromatic images at 2300~meV (in green) and
1980~meV (in red to orange) superimposed on the SEM image; (b) CL
spectrum ; (c) intensity profile of the QD emission.} \label{fig2}
\end{figure}

Figure \ref{fig2} summarizes the CL results obtained on the same NW.
Two features appear on the monochromatic images (a) and on the
spectra (b): a broad structured band around 2300~meV, visible over
the whole NW, and an instrumentally-broadened line at 1980~nm,
detected only at the tip. The high-energy side of the green band
coincides with the emission of ZnTe NWs with a (Zn,Mg)Te shell,
\emph{i.e.}, the ZnTe exciton shifted by the strain induced by the
(Zn,Mg)Te shell \cite{Artioli2013,Wojnar2014,David2014}. In
addition, a low-energy tail is observed in Fig.~\ref{fig2}b, that we
ascribe to the presence of the shallow (Zn,Cd)Te internal shell.
Such a tail, and structures at even lower energy, are observed in
other NWs with CdTe insertions, and similar features have been
identified as due to lateral QDs \cite{Wojnar2016}. They can be
efficiently eliminated by using adapted growth conditions
\cite{Orru}.

The red emission is coming from a small region, located around the
(Cd,Mn)Te insertion identified by EDX. The intensity is plotted in
Fig.~\ref{fig2}c as a function of the electron beam position. The
active region is broader than the excitation volume (calculated to
be 100~nm wide using the CASINO software\cite{Casino}), showing that
the electrons diffuse towards the QD over a typical diffusion length
which we estimate to be $L_D\approx$120~nm on both sides.

\begin{figure}
\centering
\includegraphics[scale=1]{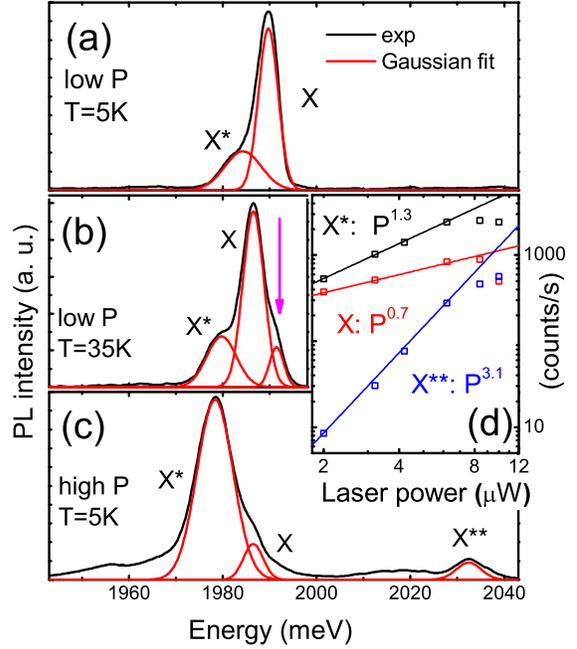}
\caption{(a) to (c) PL spectra: (a) at low temperature and very low excitation density, (b) at higher temperature, the arrow
indicates the line visible at high temperature and attributed to an excited state (c) intensity of the main lines \emph{vs.} the excitation density;(d) PL spectrum at higher excitation density and low temperature.} \label{fig4}
\end{figure}

Complementary information on this red emission is obtained by
micro-PL, Fig.~\ref{fig4}, with a better spectral resolution. At low
excitation density and low temperature, Fig.~\ref{fig4}a, the spectrum is dominated by a
line (labeled $X$) at 1989.6~meV, with a FWHM of about 5.5~meV; a
weak satellite is observed on the low energy side at 1984.2~meV
(labeled $X^*$), with a FWHM of 9~meV. This satellite strongly gains
in intensity upon increasing the excitation density
(Fig.~\ref{fig4}c), and other lines appear (for instance, $X^{**}$
at 2040~meV). The three lines exhibit a markedly different power
dependence, Fig.~\ref{fig4}d. The slightly sublinear increase and
saturation of line $X$ at 1989.6~meV suggests a neutral exciton
state. The slightly superlinear increase of the satellite $X^*$
suggests a charged exciton or possibly a bi-exciton. The third line
($X^{**}$ at 2040~meV) exhibits a strongly super-linear dependence
which suggests a multi-excitonic state. Finally, at higher
temperature, another line appears on the high-energy side (violet
arrow in Fig.~\ref{fig4}b) while the ratio between the two previous
ones is not changed significantly: the new line will be attributed
below to an excited state.

%insertion figure Fit

\begin{figure*}
\centering
\includegraphics[scale=1]{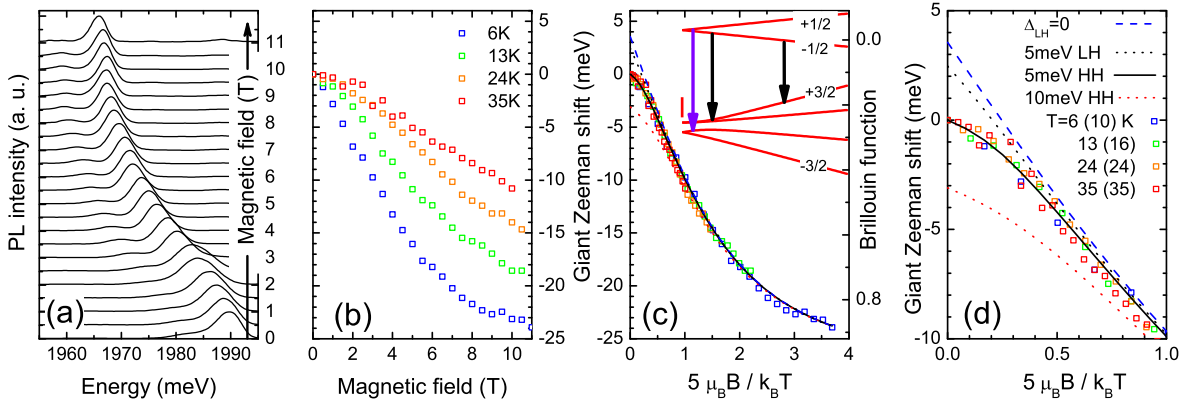}
\caption{(a) PL spectra of the QD at very low excitation density, with a magnetic field applied
perpendicular to the NW axis; (b) X line energy shift as a function
of the applied magnetic field for different values of the temperature as indicated; (c) and (d) same data, as a function of  $B/T$  (the
Mn temperature used for the plot is indicated within parentheses).
The solid line shows the final fit as described in the text, and the
dashed or dotted lines the best fit of the high-field data assuming other
values of the light-hole / heavy-hole splitting $\Delta_{LH}$. At zero splitting,
the shift is proportional to the Brillouin function (blue dashed line with the left axis for the energy shift and the
right-hand scale in (c) for the Brillouin function). The dotted lines assume two other values of $\Delta_{LH}$, 5 meV in favour of the light hole (black, upper dotted line) or 10 meV in favour of the heavy hole (red, lower dotted line). Inset of (c): scheme of the electron and hole levels,
identifying the two transitions, using the best fit parameters (that of the solid line in the main figure). The red vertical bar corresponds to a hole (electron) splitting of 10 meV, the horizontal scale is the magnetization from 0 to saturation. The state labelling is relevant at large magnetization.} \label{fig5}
\end{figure*}

Photoluminescence spectra with a magnetic field $B$ applied
perpendicular to the NW axis are shown in Fig.~\ref{fig5}a: the $X$
line exhibits a strong red shift, up to 24~meV at 11~T and 6~K. The
shift decreases as the temperature $T$ increases, as shown by the plot in Fig.~\ref{fig5}b. This temperature-dependant redshift is a characteristic
feature of the giant Zeeman effect in a diluted magnetic
semiconductor such as (Cd, Mn)Te \cite{Gaj1}: in the absence of anisotropy, the
shift is expected to be proportional to the magnetization of the Mn
spins, and hence to vary as a modified Brillouin function
$\textrm{B}_{5/2}[g\mu_BB/k_B(T+T_{AF})]$, where $g=2.0$ is the
Land\'{e} factor of Mn, $\mu_B$ is the Bohr magneton and $k_B$ the
Boltzmann constant, and $T_{AF}$ is a characteristic temperature
determined by the Mn content of the alloy \cite{Gaj1, Gaj2}.
Fig.~\ref{fig5}c shows that the redshift is indeed a saturating
function of $B/(T+T_{AF})$ (provided we accept, as usual for PL
spectra, a small heating of the Mn temperature, as indicated within
parentheses). An example of the Brillouin function is shown by the
blue dashed line: the agreement is good at high field, confirming
that the saturation of the redshift reflects that of the Mn
magnetization. There is however a clear deviation at low field, see Fig.~\ref{fig5}d: we
will show below that this deviation is due to the spin anisotropy of
the holes, and that it agrees with a heavy-hole character.

The energy of the photoluminescence line labeled X, and its
dependence on the excitation density, appear to be quite similar to
what is currently observed in self-assembled SK QDs made of the same
materials \cite{Clement}. In the present study, two correlated
experiments support this assignment to the (Cd,Mn)Te QD: (1) the
spatial localization of the CL emission coincides with the position
of the QD as determined by EDX, (2) the presence of a large giant
Zeeman effect which is expected from the presence of about 10\% Mn
in the QD as detected by EDX.

Note that the linewidth - much larger than in an usual, non-magnetic
QD, including SK CdTe QD in ZnTe - is expected for a (Cd,Mn)Te QD
and indeed experimentally observed in SK QDs \cite{Klopo2011} or QDs
in NWs \cite{Szymura}. It however makes it impossible to check an
antibunching effect (single photon emission) by a measure of
correlations, such as for instance in CdSe QDs in ZnSe NWs
\cite{Boun12} or CdTe QDs in ZnTe NWs \cite{Wojnar2011}.

\section{Discussion}
Coming back to the low-field behavior, we first use the simplest
description of a (Cd,Mn)Te QD as initially developed for SK QD
\cite{Klopo2011} and recently used for QDs in NWs \cite{Szymura}: it
involves, in the valence band, a heavy-hole ground state and a
light-hole excited state, with a splitting $\Delta_{LH}$, and a
giant Zeeman effect. Typical values of the splitting are
$\Delta_{LH}>30$~meV in SK QDs \cite{Besombes} and $\Delta_{LH}$=10
to 60~meV in QDs in NWs \cite{Szymura}. The giant Zeeman effect in
bulk material is described by an exchange field parallel and
proportional to the magnetization \textbf{M} of the Mn spins, with a
Hamiltonian \cite{Gaj1} $2Z_{hh}\textbf{s}_h.\textbf{M}/M_{sat}$
where $\textbf{s}_h$ is the hole spin operator ($=\textbf{J}_h/3$)
and the magnetization is described by a Brillouin function
$M/M_{sat}=\textrm{B}_{5/2}[g\mu_BB/k_B(T+T_{AF})]$, see above;
$Z_{hh}$ is the Zeeman shift for Mn magnetization at saturation
($M=M_{sat}$, Brillouin function equal to unity). The electron state
is described similarly by a giant Zeeman term
$2Z_{e}\textbf{s}_e.\textbf{M}/M_{sat}$. Experimentally, the maximum
giant Zeeman shifts are observed for a uniform Mn content
$x\approx0.1$, and amount to $Z_e=11$~meV for the electrons, and (if
the magnetization is along the quantization axis) $Z_{hh}=44$~meV
for the heavy hole and $Z_{lh}=44/3$~meV for the light hole
\cite{Gaj2}. The simplest adaptation to the case of QDs, the
so-called exchange box model \cite{GGG} gives similar expressions.

The solid line in Fig.~\ref{fig5}c-d shows that a good fit of the
shift of the ground-state neutral exciton and the excited state
(observed at higher temperature), as a function of the applied field
and for different values of the temperature, is obtained for the
following values of the parameters: $\Delta_{LH}=5$~meV,
$Z_e=8.5$~meV, $Z_{hh}=22.5$~meV. A clear deviation is observed at
low field if we fit the high-field data assuming a
degenerate hole multiplet ($\Delta_{LH}=0$): then the shift is simply proportional to the Brillouin
function (upper, blue dashed line in Fig.~\ref{fig5}c-d); a deviation is also observed if we assume a heavy-hole ground state with a larger splitting
$\Delta_{LH}=10$~meV (lower, red dotted line), or a light-hole
ground state (upper, black dotted line).

Let us discuss the values of the fitting parameters. The value of $\Delta_{LH}$ nicely matches the position of the
excited line observed at high temperature (violet arrow in
Fig.~\ref{fig4}b at 35K, and in the inset of Fig.~\ref{fig5}c). The
values of the giant Zeeman splitting are smaller than expected for the Mn
content $x\approx0.1$ measured by EDX: the reduction is small for the electron ($Z_e=8.5$~meV instead of 11 meV); it is larger for the heavy hole, $Z_{hh}=22.5$~meV instead of 44 meV, definitely beyond the error bar on $x$. In addition, assuming a different value of
the Mn content would affect both electrons and
holes, hence it is ruled out. Such reduced values of the giant Zeeman
splitting suggest that the electron envelope function, and even more
that of the hole, are not strictly confined to the area of the Mn
distribution. In this case, the Mn magnetization enters the
expression of the giant Zeeman effect, with a weight given by the
squared envelope function of the carrier: this causes a significant
reduction if the carrier is not well confined in the QD. In the
present case, the valence band offset between CdTe and ZnTe is small
\cite{Spaeth, Calatayud}, so that the holes are likely to be at
least partly delocalized. Indeed, long lifetimes, in the ns range,
have been measured in other QDs from the same sample, which also
suggests a non total overlap of the electron and hole.

\begin{figure}
\centering
\includegraphics*[scale=1]{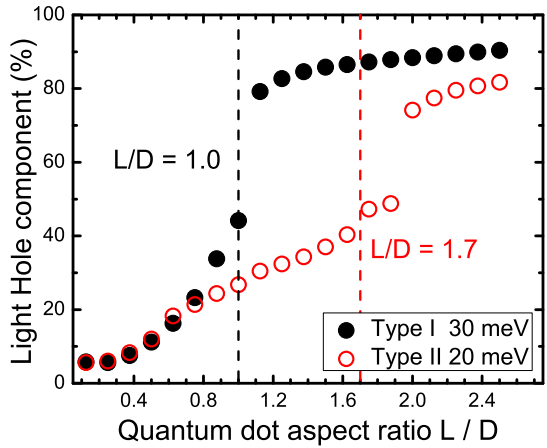}
\caption{Weight of the light-hole component for an ellipsoidal
inclusion in a ZnTe core with an outer (Zn,Mg)Te shell. Closed black
circles correspond to the well-confined state which are calculated
for a small positive value of the band offset. Open red circles
correspond to a small negative value, with a still confined state
for small values of the aspect ratio, and a multiplet at the QD-core
interface for large values of the aspect ratio. } \label{fig6}
\end{figure}

The most noticeable result is that the ground state is a heavy hole,
with a very small splitting, 5~meV, from a light-hole excited state.
This is smaller than the values of the splitting (10 to 60~meV)
previously reported \cite{Szymura} for inclusions of unknown shape.
A heavy-hole to light-hole crossing is expected when increasing the
aspect ratio of a QD across a value close to unity for QDs with
abrupt interfaces and strong valence band offset
\cite{Niquet,Zielinski}, confirming the simple idea obtained for the
ellipsoidal inclusion of an isotropic material in an infinite matrix
with a smaller lattice parameter: the crossing is predicted at unity
(sphere), both from the sign of the built-in axial shear strain
(Eshelby model \cite{Eshelby}) and from the confinement. Indeed,
light-holes excitons have been confirmed in several other QDs from
the same sample \cite{Jeannin} as the present one. In the present
QD, the crossing is not far but not reached yet, although the EDX
profile unambiguously demonstrates an elongated shape.

To understand the origin of a shift of the heavy-hole to light-hole
crossing to values of the aspect ratio larger than one, we have
performed a numerical calculation of the hole state in an
ellipsoidal QD. The calculation takes into account the elastic
coefficients, Bir-Pikus parameters, Luttinger parameters and
piezoelectric coefficient of CdTe in the inclusion, and those of
ZnTe in the ZnTe core and the Zn$_{0.9}$Mg$_{0.1}$Te outer shell. A
critical parameter is the band offset between the inclusion material
and the core material, which is known to be small for CdTe-ZnTe:
photoelectron spectroscopy studies \cite{Spaeth} conclude to a
valence band offset between 50~meV (type I) and -100~meV (type II),
while the optical spectroscopy of CdTe-(Cd,Zn)Te quantum wells
\cite{Calatayud} suggests a valence band offset equal to $\pm
50$~meV, strongly affected by the mismatch strain. Results of the
calculation are displayed in Fig.~\ref{fig6} for a QD in a
core-shell NW, for two values of the valence band offset within this
range. This - and other results \cite{Moratis} not shown  - show
that the crossover from heavy-hole to light-hole remains close to
$L/D=1$, even for a small value of the valence band offset, provided
the hole remains confined to the QD. If the valence band offset
further decreases (red open circles in Fig.~\ref{fig6}), the hole
leaks into the core for values of $L/D$ around unity, and the
mismatch with the external shell enhances the heavy-hole
contribution, thus pushing the crossover to higher values of $L/D$.
As the introduction of 10\% Mn in the CdTe inclusion increases the
bandgap by $\sim150$~meV, with about 1/3 in the valence band, the
present sample is likely to correspond to this slightly type-II
configuration, in agreement with the observation of a heavy-hole
exciton with a long lifetime for an elongated inclusion. A CdTe
inclusion possibly realizes the QD configuration represented by
black solid circles in Fig.~\ref{fig6}. To achieve the same
configuration with a (Cd,Mn)Te QD, the inclusion must be fully
embedded in a (Zn,Mg)Te NW with no ZnTe core \cite{Moratis}. A
definitive conclusion calls for a systematic study with samples on
both sides of the crossover, and the extension of the numerical
calculation to address more complex QD morphologies and to
incorporate excitonic effects.

\section{Conclusion}

To sum up, we have demonstrated that using a dedicated TEM grid / PL
sample holder based on a silicon nitride membrane allows us to apply
a broad range of experimental methods to the same, single
nano-object. Here, scanning electron microscopy with low-temperature
cathodoluminescence, transmission electron microscopy with
energy-dispersive X-ray spectrometry, and low-temperature
micro-photoluminescence with applied magnetic field, were applied to
the same, single NW containing a QD.

EDX confirms the elongated shape of the inclusion, $\sim 8$~nm $\times$ 12~nm, and its composition, Cd$_{0.9}$Mn$_{0.1}$Te. CL confirms the attribution of an emission line at $\sim2.0$ eV to this inclusion, while another line at $\sim2.3$ eV is due to the ZnTe shell.  Magneto-optical spectroscopy
identifies the heavy-hole character of the ground state exciton, with a sizable giant Zeeman effect. A quantitative study of the Zeeman effect suggests a significant delocalization of the heavy hole from the inclusion, and the presence of the light-hole exciton $\sim5$~meV higher in energy, in agreement with the onset of a PL line when increasing the temperature from 6K to 35K. Finally, a numerical calculation underlines the role of the valence band offset and the effect of a mismatched shell in setting the quantum dot aspect ratio at which the crossover from heavy-hole to light-hole takes place: the valence band offset between CdTe and ZnTe, known to be small, is further reduced by the inclusion of Mn in CdTe, pushing the (Cd,Mn)Te in ZnTe QD towards a slightly type-II configuration; then the presence of a Zn$_{0.9}$Mg$_{0.1}$Te adds a strain tensile along the NW axis and gives a heavy-hole character to the hole ground state.

\ack {This work was performed in the joint CNRS-CEA group
"Nanophysique \& semiconducteurs", the team "Laboratory of Material
Study by Advanced Microscopy", and the team "Materials, Radiations,
Structure". We acknowledge funding by the French National Research
Agency (project Magwires ANR-11-BS10-013, COSMOS ANR-12-JS10-0002,
and ESPADON ANR-15-CE24-0029).}

%\Bibliography{<10>}

%\endbib

\end{document}